\documentclass[preprint,aps,showpacs,nofootinbib]{revtex4}
\usepackage{mathrsfs}
\usepackage{amsmath}
\usepackage{graphicx}
\usepackage{subfigure}
\usepackage{bm}
\usepackage{epstopdf}

\DeclareMathAlphabet\mathbfcal{OMS}{cmsy}{b}{n}
\newcommand{\bea}{\begin{eqnarray}}
\newcommand{\eea}{\end{eqnarray}}
\newcommand{\beq}{\begin{equation}}
\newcommand{\eeq}{\end{equation}}
\newcommand{\nn}{\nonumber}

\def\/{\over}
\begin{document}
\title{\bf Casimir-Polder force for a polarizable molecule near a dielectric substrate out of thermal equilibrium}
\author{Wenting Zhou$^1$ and Hongwei Yu$^{1,2}$}
\affiliation{$^1$ Center for Nonlinear Science and Department of Physics, Ningbo
University, Ningbo, Zhejiang 315211, China\\
$^{2}$ Institute of Physics and Key Laboratory of Low
Dimensional Quantum Structures and Quantum
Control of Ministry of Education,\\
Hunan Normal University, Changsha, Hunan 410081, China }

%\affiliation{Center for Nonlinear Science and Department of Physics, Ningbo
%University, Ningbo, Zhejiang 315211, China}

\begin{abstract}

We demonstrate that the Casimir-Polder force for a  molecule  near the surface of a real dielectric substrate  out of thermal equilibrium displays distinctive behaviors as compared to that at thermal equilibrium. In particular, when the molecule-substrate separation is much less than the molecular transition wave-length, the CP force in the high temperature limit can be dramatically manipulated by varying the relative magnitude of the temperatures of the substrate and the environment so that the attractive-to-repulsive transition can occur beyond certain a threshold temperature of either the substrate or the environment  depending on which one is higher for molecules both in the ground and excited states. More remarkably, when the separation is comparable to the wave-length, such transitions  which are impossible at thermal equilibrium may happen for longitudinally polarizable molecules  with a small permittivity, while for isotropically polarizable ones the transitions can even occur  at   room temperature for some dielectric substrates such as sapphire and graphite which is much lower than the temperature for the transition to take place in the thermal equilibrium case, thus making the experimental demonstration of such force transitions easier.
\end{abstract}
\pacs{31.30.jh, 12.20.-m, 34.35.+a, 42.50.Nn} \maketitle

%\section{Introduction}

Casimir and Polder  demonstrated in 1948 the existence of a force between an atom and a conducting plate~\cite{C-P48}, which is what we now call the Casimir-Polder (CP) force. The CP force is a typical quantum effect arising from the interaction between the atom and  the zero-point  fluctuations of quantum electromagnetic fields. Subsequent studies show that besides zero-point fluctuations, thermal fluctuations also contribute to the atom-wall force~\cite{Lifshitz55,Lifshitz61}. On the one hand, the CP force is  a fascinating topic  in fundamental research, and on the other hand,  it is becoming increasingly important  in technological applications~\cite{Craighead00,Sugimoto07,Chan,NT05}
%and in biological systems~\cite{Gingell} ,
and  relevant in trapping and coherently manipulating cold atoms and even polar molecules near surfaces~\cite{Lin04,VBM05}.  Although the force is usually very tiny, it has been measured with remarkable precision~\cite{Sando,Sukenik,Landragin,Druzh,Oberst}.

The origin of the CP force at finite temperature is generally attributed to two independent sources, i.e., zero-point fluctuations and thermal fluctuations. At short distances, the contribution of zero-point fluctuations is proportional to $z^{-4}$ with $z$ being the distance between the atom and the wall, and it dominates over the contribution of thermal fluctuations, while at distances much larger than the wavelength of thermal photons, $z\gg{\hbar c\/k_{B} T}$, the contribution of thermal fluctuations is proportional to $T/z^{4}$, which provides the leading contribution to the total force.

However, this behavior of the CP force changes when the situation comes to that of out of thermal equilibrium~\cite{Henkel,Cohen03,Antezza05,Antezza06,Obrecht,Dedkov,Ellingsen09}. In this regard, Antezza et al found that  for an atom at very large distances from the surface of a half-space dielectric substrate, the CP force exhibits, when the temperature is low, a new, stronger asymptotic behavior out of thermal equilibrium as compared to that in the equilibrium case~\cite{Antezza05,Antezza06}, which has been verified in experiment~\cite{Obrecht}~\footnote{Let us  note here that the thermal CP force at thermal equilibrium has also been experimentally observed recently~\cite{Sushkov}.}, providing the first experimental observation of the thermal effect in the CP force. Later, Dedkov and Kyasov extended the analysis to different nonequilibrium configurations~\cite{Dedkov} and Ellingsen et al showed that Antezza et al's result can also be obtained in the framework  of the Keldysh Green function method~\cite{Ellingsen09}.
This has spurred a great deal of interest in the Casimir effect out of thermal equilibrium~~\cite{Buhmann092,Sherkunov09,Hu1,Hu2,Hu3,Buhmann08,Antezza06L,Antezza08,Bimonte09,Messina,Kruger,Kruger11L,Kruger12} as well as other effects such as manipulation of atomic populations~\cite{Bellomo612,Bellomo113} and entanglement~\cite{Bellomo132} in situations out of thermal equilibrium. Recently, by generalizing the formalism proposed by Dalibard, Dupont-Roc and Cohen Tannoudji~\cite{DDC82,DDC84} to the case out of thermal equilibrium, we studied in detail the out-of-thermal-equilibrium CP force of an atom (both in the ground and excited states) near a half-space dielectric substrate in the short, intermediate and long distance regions in both the low- and the high-temperature limits~\cite{Z-Y14}. Our result  on the CP force in the long distance region and low-temperature limit recovers that of  M. Antezza  et al and furthermore we quantify the meaning of ``very large distances" which is taken as a mathematical infinity ($z\rightarrow\infty$) in \cite{Antezza05,Antezza06} by giving a concrete region where this behavior holds~\cite{Z-Y14}.

In the present paper, we investigate the CP force of a typical polarizable molecule with long-wavelength transitions near the surface of a dielectric substrate out of thermal equilibrium. At this point, let us note that Ellingsen, et. al demonstrated that for a typical molecule with long-wavelength transitions placed near a plane metal surface at thermal equilibrium, the CP potentials can be entirely independent of temperature even when the thermal photon number is large~\cite{Ellingsen}, while
we showed that when the molecular transition wavelengths are comparable to the molecule-surface separation, the CP force can be dependent on the ambient temperature and the molecular polarization, and it can even change from attractive to repulsive at room temperature~\cite{Z-Y12}. What we are particularly interested in here are the new features coming from being out of thermal equilibrium with a real dielectric substrate  as opposed to thermal equilibrium and a metal surface.

We assume, for simplicity, that the dielectric substrate is non-dispersive and the molecule is modeled by a two-level system with two stationary states represented by $|+\rangle$ and $|-\rangle$, and energy spacing between the two levels being $\hbar\omega_{0}$. The left half-space with $z<0$ is occupied by a dielectric substrate with real relative permittivity $\epsilon$ and temperature $T_{s}$ with the surface of the substrate coinciding with the plane $z=0$, and the right half-space with $z>0$ is occupied by a thermal bath (environment) with temperature $T_{e}$. Generally, the temperatures of the substrate and the environment do not coincide and each half-space is assumed to be in local thermal equilibrium. The molecule is placed at a distance $z>0$ from the surface of the substrate. We define the polarizability of the molecule in an arbitrary state $|a\rangle$ as
\beq
\alpha=\sum_{i}\alpha_{i}=\sum_{i,b}{2|\langle a|\mu_{i}(0)|b\rangle|^{2}\/{3\hbar\omega_{0}}}
\eeq
where $\mu_{i}$ is the $i-th$ spatial component of the dipole moment of the particle, and the summation over $b$ extends over its complete set of states.

The  position-dependent particle-surface potential for the state $|a\rangle$ can be expressed in  a sum of three parts as~\cite{Z-Y14},
\beq
(U_{a})^{bnd}_{tot}(z)=(U_{a})^{bnd}_{vac}(z)+(U_{a})^{bnd}_{eq}(z,\beta_{e})+(U_{a})^{bnd}_{neq}(z,\beta_{s},\beta_{e})
\label{general E tot}
\eeq
with $\beta_{e}={\hbar c\/{k_{B}T_{e}}}$ and $\beta_{s}={\hbar c\/{k_{B}T_{s}}}$ being the wavelength of thermal photons, $(U_{a})^{bnd}_{vac}(z)$ representing the contribution of zero-point fluctuations, $(U_{a})^{bnd}_{eq}(z,\beta_{e})$  the contribution of thermal fluctuations at thermal equilibrium at a temperature $T_{e}$, and $(U_{a})^{bnd}_{neq}(z,\beta_{s},\beta_{e})$  the contribution of non-thermal equilibrium. The explicit expressions for the above three parts are
\bea
(U_{a})^{bnd}_{vac}(z)&=&-{3\hbar\omega_{0}\/{4\pi^{2}\varepsilon_{0}c^{3}}}
\int^{\infty}_{0}d\omega {\omega^{3}\/{\omega-\omega_{ab}}}\sum_{\sigma}\alpha_{\sigma} f_{\sigma}(z,\omega)\;,
\label{E vac for real dielectric}\\
(U_{a})^{bnd}_{eq}(z,\beta_{e})&=&{3\hbar\omega_{0}\/{4\pi^{2}\varepsilon_{0}c^{3}}}
\int^{\infty}_{0}d\omega\biggl({\omega^{3}\/{\omega+\omega_{ab}}}-{\omega^{3}\/{\omega-\omega_{ab}}}\biggr){1\/{e^{\beta_{e}\omega/c}-1}}
\sum_{\sigma}\alpha_{\sigma} f_{\sigma}(z,\omega)\;,
\label{E eq for real dielectric}\nn\\\\
(U_{a})^{bnd}_{neq}(z,\beta_{s},\beta_{e})&=&{3\hbar\omega_{0}\/{4\pi^{2}\varepsilon_{0}c^{3}}}
\int^{\infty}_{0}d\omega\biggl({\omega^{3}\/{\omega+\omega_{ab}}}-{\omega^{3}\/{\omega-\omega_{ab}}}\biggr)
\biggl({1\/{e^{\beta_{s}\omega/c}-1}}-{1\/{e^{\beta_{e}\omega/c}-1}}\biggr)\nn\\
&&\quad\quad\times\sum_{\sigma}\alpha_\sigma\int^{1}_{0}dt\; A_\sigma(t)e^{-2z\sqrt{\epsilon-1}\omega t/c}
\label{E neq for real dielectric}
\eea
where $\sigma=\parallel,\perp$, $\varepsilon_{0}$ is the permittivity of vacuum, $\alpha_{\parallel}=\alpha_{x}+\alpha_{y}$, $\alpha_{\perp}=\alpha_{z}$ and
\beq
f_{\sigma}(z,\omega)=\int^{1}_{0}dt\;[A_{\sigma}(t)e^{-2z\sqrt{\epsilon-1}\omega t/c}+T_{\sigma}(t)\cos(2z\omega t/c)]\;.\label{function f}
\eeq
In the above equations, we have defined
\bea
A_{\parallel}(t)&=&{1\/2}\sqrt{\epsilon-1}{{(2\epsilon+1)(\epsilon-1)t^{2}+1}\/{(\epsilon^{2}-1)t^{2}+1}}t\sqrt{1-t^{2}}\;,\label{concrete A11}\\
A_{\perp}(t)&=&\epsilon\sqrt{\epsilon-1}{{(\epsilon-1)t^{2}+1}\/{(\epsilon^{2}-1)t^{2}+1}}t\sqrt{1-t^{2}}\;,\label{concrete A1}\\
T_{\parallel}(t)&=&{1\/4}\biggl({{t-\sqrt{\epsilon-1+t^{2}}}\/{t+\sqrt{\epsilon-1+t^{2}}}}-t^{2}{{\epsilon t-\sqrt{\epsilon-1+t^{2}}}\/{\epsilon t+\sqrt{\epsilon-1+t^{2}}}}\biggr)\;,\\
T_{\perp}(t)&=&{1\/2}(1-t^{2}){{\epsilon t-\sqrt{\epsilon-1+t^{2}}}\/{\epsilon t+\sqrt{\epsilon-1+t^{2}}}}\;.\label{concrete T1}
\eea
The CP force on the particle can be obtained by taking the derivative of $z$ on the particle-surface potential, Eq.~(\ref{general E tot}),
%\beq
$F_{a}=-{\partial\/\partial z}(U_{a})^{bnd}_{tot}(z)\;.$
%\label{CP force}
%\eeq

Let us first study the temperature dependence of the CP force for a typical  molecule whose transition wavelength is much larger than the typical experimental molecule-surface separation placed near the surface of a substrate with a real relative permittivity $\epsilon$. For this purpose, we can define a geometric temperature, $T_{z}={\hbar c\/z k_{B}}$, i.e., the temperature of radiation whose wavelength is of order $z$, and a spectroscopic temperature, $T_{\omega_{0}}={\hbar \omega_{0}\/k_{B}}$, which is roughly the temperature required to noticeably populate the upper level. For the case we are considering, we have $\{{z\omega_{0}\/c},{z\sqrt{\epsilon-1}\omega_{0}\/c}\}\ll1$~\footnote{Hereafter, $\{a,b\}\gg c$ means $a\gg c$ and $b\gg c$. Similarly, $\{a,b\}\ll c$ means $a\ll c$ and $b\ll c$.},  i.e., $\{T_{z},{T_{z}\/\sqrt{\epsilon-1}}\}\gg T_{\omega_{0}}$, as long as $\epsilon$ is not very large. Then the analysis of the temperature dependence of the CP force can be divided into the following three typical regions: the low temperature region where both $T_{s}$ and $T_{e}$ are much smaller than the spectroscopic temperature, i.e., $\{T_{s},T_{e}\}\ll T_{\omega_{0}}\ll\{T_{z},{T_{z}\/\sqrt{\epsilon-1}}\}$, the intermediate temperature region where both $T_{s}$ and $T_{e}$ are much smaller than the geometric temperature and much larger than the spectroscopic temperature, i.e., $T_{\omega_{0}}\ll\{T_{s},T_{e}\} \ll \{T_{z},{T_{z}\/\sqrt{\epsilon-1}}\}$, and the high temperature region where both $T_{s}$ and $T_{e}$ are much larger than the geometric temperature, i.e., $T_{\omega_{0}}\ll \{T_{z},{T_{z}\/\sqrt{\epsilon-1}}\}\ll \{T_{s},T_{e}\}$.

Generally, the thermal radiation that originates both from the substrate and the environment contributes to the molecular CP force. In the low and intermediate temperature regions, though the contribution of thermal radiation that originates from the substrate is much larger than the contribution of radiation from the environment if $T_{s}/T_{e}$ is not extremely small, it is much smaller than the contribution of zero-point fluctuations, and thus in these two regions, the CP force behaves like $z^{-4}$, i.e., it obeys   van-der Waals law. So, the CP force here is essentially temperature-independent just as the CP force of a molecule located near a perfect conducting plate~\cite{Z-Y12}.

However, if we go to the high temperature region, $T_{\omega_{0}}\ll \{T_{z},{T_{z}\/\sqrt{\epsilon-1}}\}\ll\{T_{s},T_{e}\}$, i.e., $\{\beta_{s},\beta_{e}\}\ll \{z,z\sqrt{\epsilon-1}\}\ll\lambda_{0}$ where $\lambda_{0}={c\/\omega_{0}}$ is the wavelength of the molecule, then dependence shows up and the CP force of the ground and excited molecules near the substrate out of thermal equilibrium can now be written respectively as
\bea
F_{-}&\approx&-{\hbar\/{4\pi\varepsilon_{0}}}\biggl\{{{\epsilon-1}\/{\epsilon+1}}{9\omega_{0}(\alpha_{\parallel}+2\alpha_{z})\/{32z^{4}}}
+\biggl({\alpha_{\parallel}\/2}+\alpha_{z}\biggr){3\omega_{0}^{2}\/4z^{2}c\beta_{e}}\nn\\&&\quad\;\quad\;\quad\;
-\biggl[{\alpha_{\parallel}\/2}g_{1}(\epsilon)+\alpha_{z}g_{2}(\epsilon)\biggr]{3\omega_{0}^{2}\/4z^{2}c\beta_{s}}\biggr\}\;,\label{high temp moderate distance-ground state}\\
F_{+}&\approx&-{\hbar\/{4\pi\varepsilon_{0}}}\biggl\{{{\epsilon-1}\/{\epsilon+1}}{9\omega_{0}(\alpha_{\parallel}+2\alpha_{z})\/{32z^{4}}}
-\biggl({\alpha_{\parallel}\/2}+\alpha_{z}\biggr){3\omega_{0}^{2}\/4z^{2}c\beta_{e}}\nn\\&&\quad\;\quad\;\quad\;
+\biggl[{\alpha_{\parallel}\/2}g_{1}(\epsilon)+\alpha_{z}g_{2}(\epsilon)\biggr]{3\omega_{0}^{2}\/4z^{2}c\beta_{s}}\biggr\}\label{high temp moderate distance-excited state}
\eea
with
\beq
g_{1}(\epsilon)={{2\epsilon^{2}+\epsilon+1}\/{(\epsilon+1)^{2}}}\;,\quad\;g_{2}(\epsilon)={{(3\epsilon+1)\epsilon}\/{(\epsilon+1)^{2}}}\;.
\eeq
Let us note here that although Eqs.~(\ref{high temp moderate distance-ground state}) and (\ref{high temp moderate distance-excited state}) have well-defined perfect-reflector limits ($\epsilon\rightarrow\infty$), they are not valid for the case of a molecule located near a perfect reflector as the condition $\{z,z\sqrt{\epsilon-1}\}\ll\lambda_{0}$ breaks down. Just as has been pointed out previously~\cite{Z-Y14},
%in Ref.~\cite{Eberlein},
for the case of a perfect conducting plate, the limit $\epsilon\rightarrow\infty$ should be taken in all expressions before analyzing the asymptotic behaviors.

We now first examine the case of thermal equilibrium and see how it differs from the case of a metal plate. When $T_{s}$ coincides with $T_{e}$, i.e., $T_{s}=T_{e}=T$, the above results reduce to the total CP force for the molecule at thermal equilibrium as
\bea
F_{-}&\approx&-{\hbar\/{4\pi\varepsilon_{0}}}\biggl\{{{\epsilon-1}\/{\epsilon+1}}{9\omega_{0}(\alpha_{\parallel}+2\alpha_{z})\/{32z^{4}}}
-\biggl[{\alpha_{\parallel}\/2}g_{3}(\epsilon)
+\alpha_{z}g_{4}(\epsilon)\biggr]{3\omega_{0}^{2}\/4z^{2}c\beta}\biggr\}\;,\label{eq--}\\
F_{+}&\approx&-{\hbar\/{4\pi\varepsilon_{0}}}\biggl\{{{\epsilon-1}\/{\epsilon+1}}{9\omega_{0}(\alpha_{\parallel}+2\alpha_{z})\/{32z^{4}}}
+\biggl[{\alpha_{\parallel}\/2}g_{3}(\epsilon)
+\alpha_{z}g_{4}(\epsilon)\biggr]{3\omega_{0}^{2}\/4z^{2}c\beta}\biggr\}\label{eq-+}
\eea
with
\beq
g_{3}(\epsilon)={{\epsilon(\epsilon-1)}\/{(\epsilon+1)^{2}}}\;,\quad\;\quad\;g_{4}(\epsilon)={{(2\epsilon+1)(\epsilon-1)}\/{(\epsilon+1)^{2}}}
\eeq
and $\beta={\hbar c\/k_{B}T}$. For the ground-state molecule, there exists a threshold temperature,
\beq
T_{0}={{3\hbar c^{2}}\/{4k_{B}z^{2}\omega_{0}}}{{\alpha_{\parallel}+2\alpha_{z}}\/{\alpha_{\parallel}{\epsilon\/\epsilon+1}+2\alpha_{z}{{2\epsilon+1}\/{\epsilon+1}}}}\;,
\label{To}
\eeq
which depends on both the molecular polarization and the dielectric permittivity and beyond which the attractive-to-repulsive transition of the CP force  occurs, while for the excited molecule, the CP force is always attractive. This transition happens for molecules with any polarization. This is in sharp contrast to the case of a metal surface where the behavior of the CP force depends crucially on the sign of $\alpha_{\parallel}-2\alpha_{z}$ and the transition of attractive-to-repulsive is possible for molecules both in the ground and excited states~\cite{Z-Y12}.

When $T_{s}$ does not coincide with $T_{e}$, Eqs.~(\ref{high temp moderate distance-ground state}) and (\ref{high temp moderate distance-excited state}) describe the CP forces of a molecule  near the substrate out of thermal equilibrium. If $T_{e}\ll T_{s}$, the second term in Eqs.~(\ref{high temp moderate distance-ground state}) and (\ref{high temp moderate distance-excited state}) is much smaller than the third term, and this means that the contribution of thermal fluctuations originating from the environment is negligible as compared to that from the substrate. For the molecule in the ground state, there exists a threshold temperature of the substrate at which the first term balances the third term in  Eqs.~(\ref{high temp moderate distance-ground state}) ,
\beq
T_{s0}={{3\hbar c^{2}}\/{4k_{B}z^{2}\omega_{0}}}{{\epsilon-1}\/{\epsilon+1}}
{{\alpha_{\parallel}+2\alpha_{z}}\/{\alpha_{\parallel}g_{1}(\epsilon)+2\alpha_{z} g_{2}(\epsilon)}}\;.
\label{Tso}
\eeq
Obviously, this threshold depends again on both the molecular polarization and the permittivity of the substrate. When the temperature of the substrate is lower than that of the threshold, the CP force is attractive as a result of the fact that the first term in Eq.~(\ref{high temp moderate distance-ground state}) overtakes the third one so that $F_{-}<0$,  and repulsive otherwise, while for the molecule in the excited state,  the CP force is always attractive since $F_{+}$ is always negative  as can be seen  from Eq.~(\ref{high temp moderate distance-excited state}). This behavior is qualitatively similar to that in the case of thermal equilibrium.  However, if $T_{e}\gg T_{s}$, the second term in Eqs.~(\ref{high temp moderate distance-ground state}) and (\ref{high temp moderate distance-excited state}) is much larger than the third term, so that the contribution of thermal fluctuations that originate from the substrate is negligible as compared to that from the environment. Now for the molecule in the excited state, there exists a threshold temperature of the environment at which the first term balances the third term in Eqs.~(\ref{high temp moderate distance-excited state}) ,
\beq
T_{e0}={{3\hbar c^{2}}\/{4k_{B}z^{2}\omega_{0}}}{{\epsilon-1}\/{\epsilon+1}}\;.
\label{Teo}
\eeq
When the temperature of the environment is lower than that of the threshold, the CP force the excited molecule feels is attractive, and repulsive otherwise, while for the molecule in the ground state, the CP force is always attractive.  This indicates that the molecular CP force out of thermal equilibrium near a real dielectric substrate can be dramatically manipulated by varying the relative magnitude of the  temperatures of the substrate and the environment, and it   displays behaviors distinctive from  that at thermal equilibrium near a metal surface. Interestingly,  this threshold, unlike $T_{s0}$,  is independent of the molecular polarization.

The above analysis shows that, in the near-zone and in the high temperature limit, ($\{\beta_{s},\beta_{e}\}\ll \{z,z\sqrt{\epsilon-1}\}\ll\lambda_{0}$), the attractive-to-repulsive transition of the CP force can occur for molecules in both the ground and excited states in an environment out of thermal equilibrium under certain conditions, while for a thermal equilibrium environment, such a transition can only occur for  ground-state molecules as the CP force on the excited molecule is always attractive. Now a few comments are in order. First,  the contribution of the thermal radiation from the environment to the CP force is dependent on the molecular polarization (refer to the second terms in Eqs.~(\ref{high temp moderate distance-ground state}) and (\ref{high temp moderate distance-excited state})), while the contribution of the thermal radiation from the substrate depends on both  the molecular polarization and the dielectric properties of the substrate (see the third terms in Eqs.~(\ref{high temp moderate distance-ground state}) and (\ref{high temp moderate distance-excited state})). Moreover, they are of opposite signs, thus when the system is out of thermal equilibrium, a disparity between the contribution of the thermal radiation from the substrate and that from the environment appears when $T_{e}\ll T_{s}$ or $T_{e}\gg T_{s}$, and this  results in  a threshold temperature for either the substrate or the environment across which the attractive-to-repulsive transition of the CP force can occur for either the ground-state or excited-state molecules. However, for the thermal equilibrium case, i.e., when the temperatures of the substrate and the environment coincide, the contributions of the thermal radiation from the substrate and the environment can be incorporated into one term which gives a repulsive force component for the CP force of the ground-state molecules and an attractive one for the excited-state molecules (see Eqs.~(\ref{eq--}) and (\ref{eq-+})). As a result, at thermal equilibrium, the attractive-to-repulsive transition of the CP force can occur only for ground-state molecules while  the CP force on the excited-state molecules is always attractive in the near zone.  Second, the main interest of the present paper is the temperature dependence of the CP force on the typical long wave-length molecules. However,  one can show that the attractive-to-repulsive transition of the CP force for the excited-state molecules can also happen in the far-zone ($\{z,z\sqrt{\epsilon-1}\}\gg\lambda_{0}$) as the molecule-substrate distance varies and in fact the CP force oscillates as a function of the distance just as what happens for the CP force on an atom in excited states in the far-zone. It is worth noting however that for a typical molecule with long wave-length, the far-zone is too far for the molecular CP force to have any experimental significance.

\begin{figure}[!htb]
\centering
\subfigure[]
{\label{file11}
\includegraphics[scale=0.55]{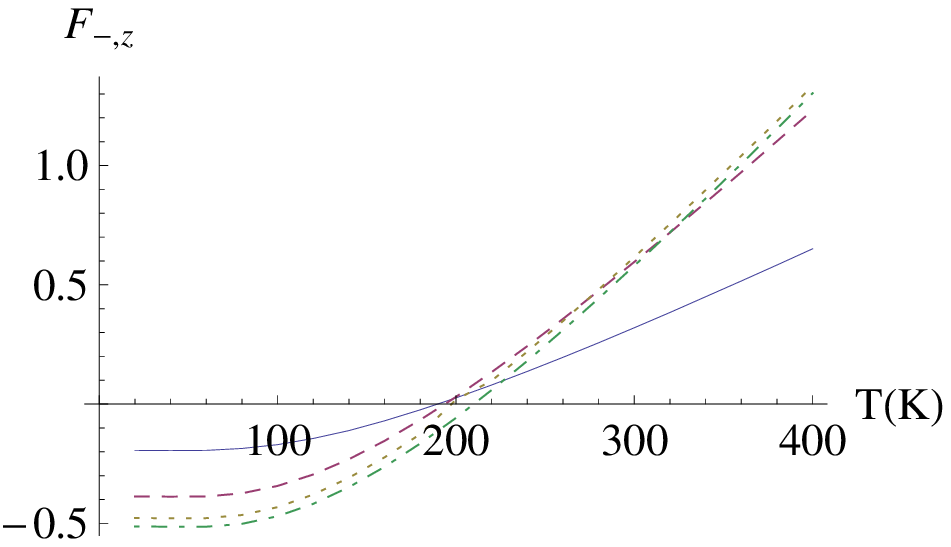}}
\subfigure[]
{\label{file21}
\includegraphics[scale=0.50]{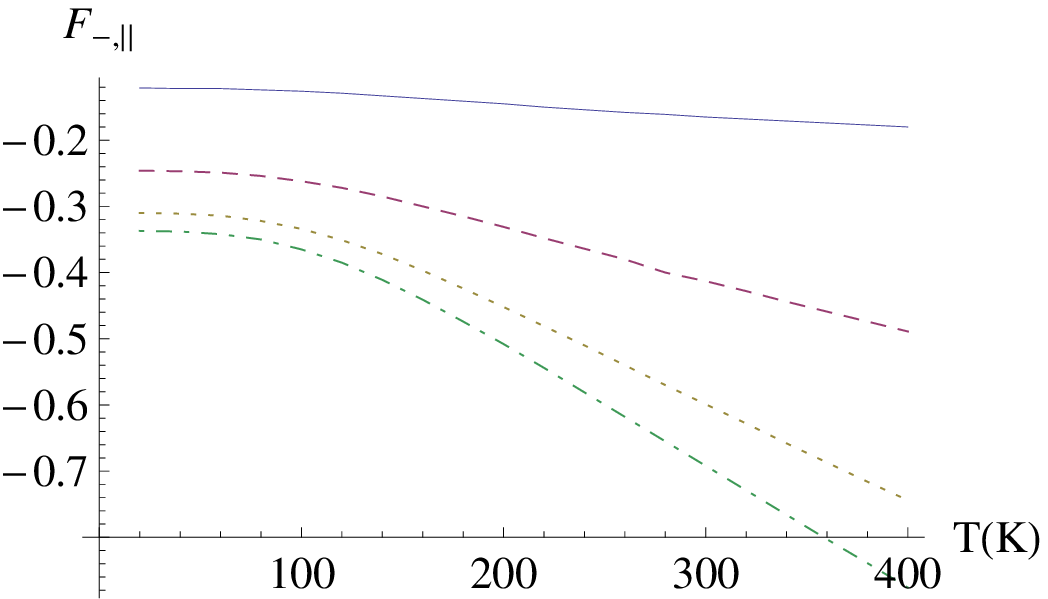}}
\subfigure[]
{\label{file31}
\includegraphics[scale=0.50]{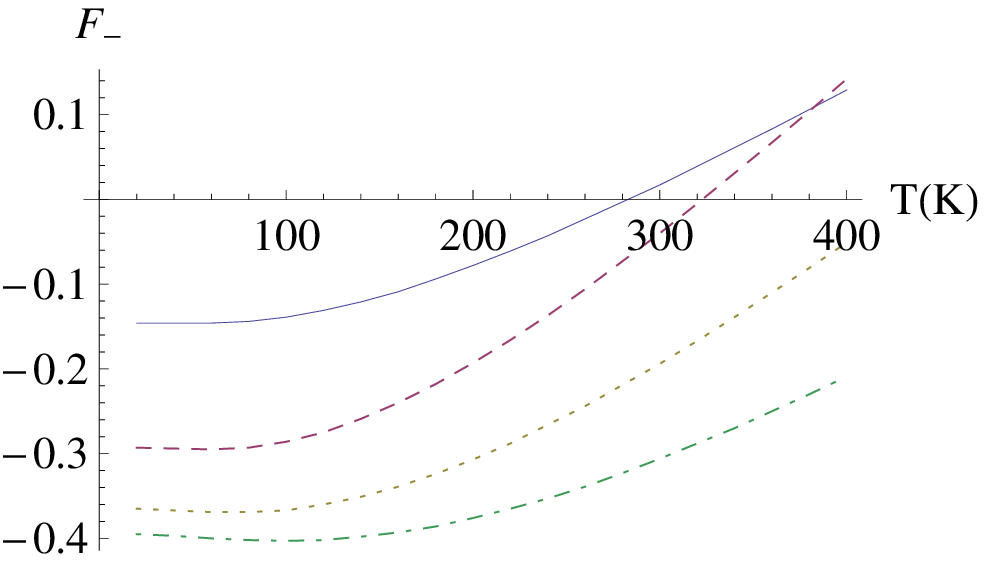}}
\caption{Casimir-Polder force for the ground-state molecule at thermal equilibrium near a real dielectric substrate when $z=6um$. The solid, dashed, dotted, dot-dashed lines correspond to $\epsilon=2,5,10,15$ respectively. The force is in the unit of ${\hbar c\omega_{0}^{2} \alpha}/(128\pi \varepsilon_{0})$. (a) The molecule is polarized along the $z$-direction; (b) parallel to the surface of the substrate; (c) isotropically.}
\end{figure}

Now we turn our attention to the CP force of the molecule at a distance comparable to its transition wave-length both in and out of thermal equilibrium near a dielectric substrate.
We start with the equilibrium case and take the LiH molecule, whose vibrational transition frequency is $\omega_{0}\sim4.21\times10^{13}\mathrm{Hz}$ ($\lambda_{0}\sim7.02\mu m$), as an example. Assume that the molecule is in the ground state and located at a distance $z=6\mu m$ from the surface of a dielectric substrate. Consider first the case in which the temperatures of the environment and the substrate coincide, $T_{s}=T_{e}$, i.e., the molecule-substrate system is at thermal equilibrium. For the ground-state molecule polarizable along the $z$-direction  and the substrate with a given value of $\epsilon$, the attractive-to-repulsive changeover of the CP force can occur at a certain temperature, and with the increasing of the value of $\epsilon$, the change occurs at higher temperature, as is shown in Fig.~\ref{file11}.
While for the molecule polarizable parallel to the surface of the substrate,  the CP force of the ground-state molecule is always attractive, as is shown in Fig.~\ref{file21}. For small $\epsilon$, the CP force varies very slowly with the temperature  in the  $T\lesssim100K$ region and is effectively temperature-independent. And in fact this temperature independence of the CP force does not change appreciably as $\epsilon$ increases. This  is similar to that of the CP force of a molecule located near a metal surface at thermal equilibrium which can be viewed as  independent of the ambient temperature though the number of photons may be large~\cite{Ellingsen,Z-Y12}.
However, in the region $T\gtrsim100K$, the CP force becomes temperature-dependent as long as $\epsilon$ is not very small, and the dependence  becomes more obvious with larger $\epsilon$. This is consistent with the result that the CP force of the same molecule near the surface of a perfect conducting plate varies dramatically with the temperature when $T\gtrsim100K$ ~\cite{Z-Y12}. For an isotropically polarizable molecule, Fig.~\ref{file31} shows that the attractive-to-repulsive changeover of the CP force can also happen, depending on the value of the relative permittivity of the substrate $\epsilon$. For the CP force of the molecule near a half-space granite substrate with $\epsilon\sim5$, the change occurs at about $320K$. However, for the same molecule located at the same distance near the surface of a perfect conducting plane, the CP force is always attractive and no attractive-to-repulsive change occurs~\cite{Z-Y12}. This demonstrates again that the behaviors of the CP force for typical long wave-length molecules  near a dielectric substrate are sharply different from those near a metal surface.

\begin{figure}[!htb]
\centering
\subfigure
{\label{file12}
\includegraphics[scale=0.50]{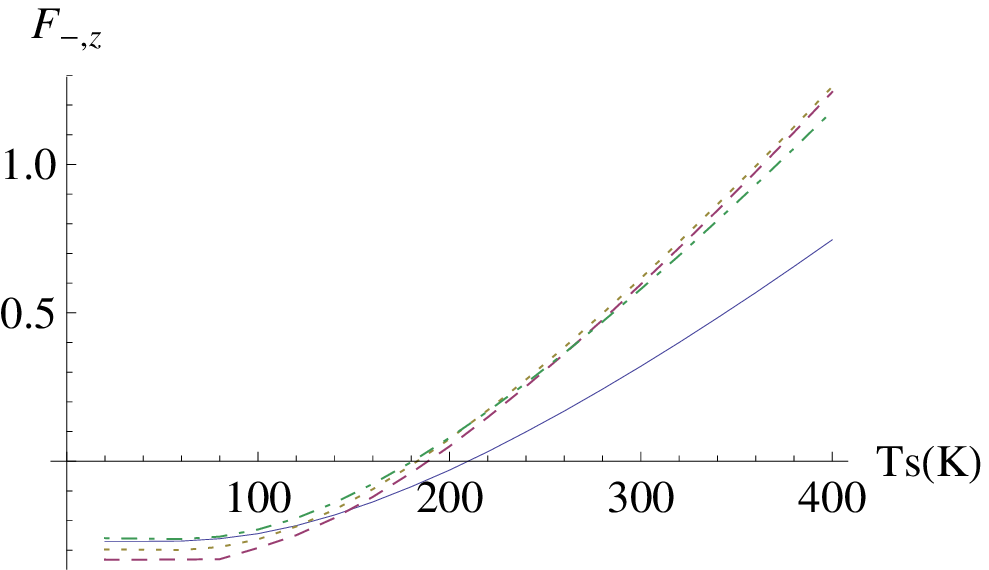}}
\subfigure[]
{\label{file22}
\includegraphics[scale=0.50]{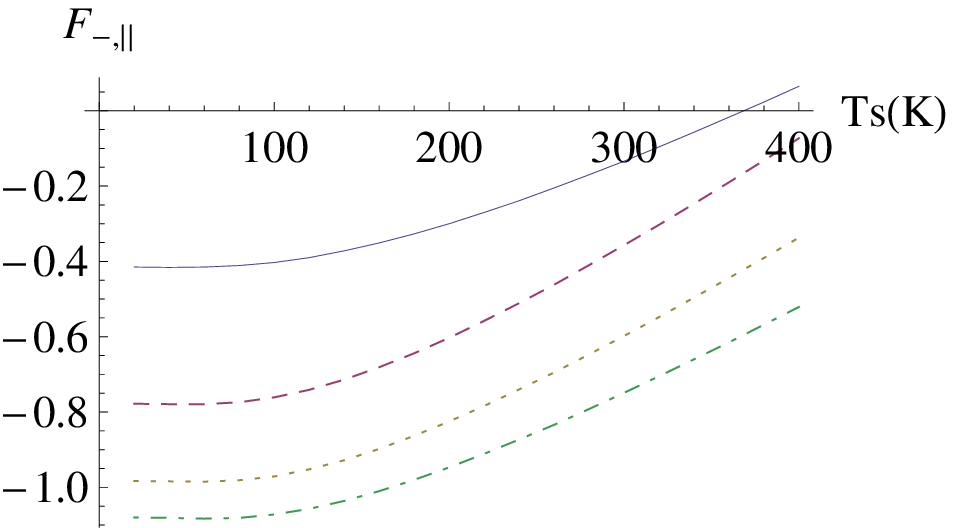}}
\subfigure[]
{\label{file32}
\includegraphics[scale=0.50]{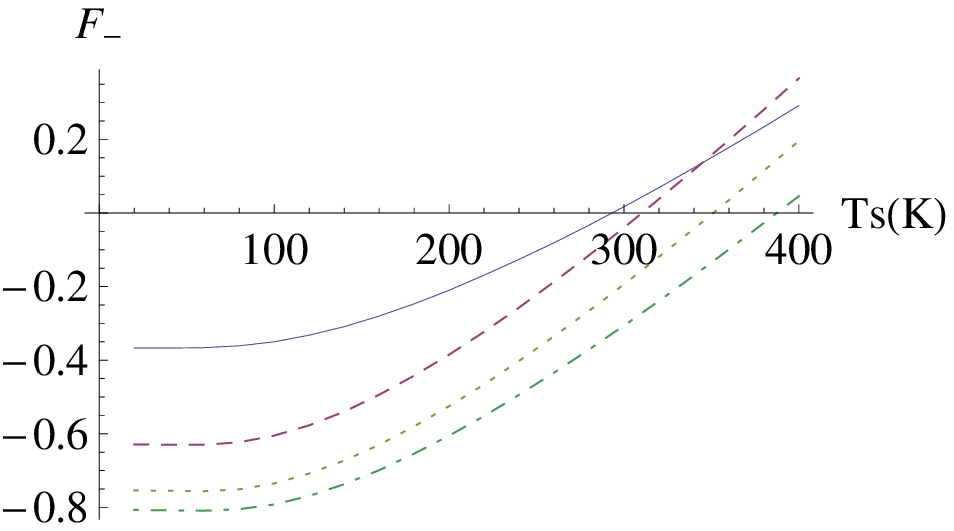}}
\caption{ Same as Fig.1, but for  the ground-state molecule out of thermal equilibrium.}
\label{3}
\end{figure}

When the temperatures of the environment and the substrate do not coincide, $T_{s}\neq T_{e}$, i.e., the system is out of thermal equilibrium. Figure~\ref{3} shows how the CP force of the ground-state molecule varies with the temperature of the substrate when the temperature of the environment is  kept at room temperature, $T_{e}\sim300K$. For transversely polarizable molecules, Fig.~\ref{file12} shows that the CP force is similar to that in the case of thermal equilibrium (see Fig.~\ref{file11}). In both cases, the attractive-to-repulsive changeover of the CP force occurs around $T_{s}\sim200K$. In contrast, for longitudinally polarizable molecules, the CP force of the molecule changes dramatically as opposed to that in the case of thermal equilibrium (see Fig.~\ref{file21}). As shown in Fig.~\ref{file22}, the attractive-to-repulsive transition of the CP force which doesn't exist in the case of thermal equilibrium (see Fig.~\ref{file21}) can occur in the out of thermal equilibrium case. Especially, for a substrate with small permittivity, a cement substrate with $\epsilon\sim2$ for example, the change occurs at $T_{s}\sim370K$. For an isotropically polarizable molecule, Fig.~\ref{file32} shows that the attractive-to-repulsive changeover of the CP force can occur when the system is out of thermal equilibrium. Especially, for a substrate with a relatively large relative permittivity, a sapphire substrate with $\epsilon\sim10$ and a graphite substrate with $\epsilon\sim15$ for example, the changes occur at $T_{s}\sim350K$ and $380K$ respectively which are much lower than the temperatures for the changeover to occur  in the case of thermal equilibrium (Fig.~\ref{file31}).

In summary, we have studied the CP force of a typical molecule with long wave-length transitions near a real dielectric substrate both in and out of thermal equilibrium.  In the case of thermal equilibrium,  we find that, when the molecule-substrate separation is much smaller than the molecular transition wave-length, the CP force, in the high temperature limit, for the molecules in excited states is always attractive while that for the ground state can change from attractive to repulsive beyond some threshold temperature  and this attractive-to-repulsive transition can happen for any molecular polarization. This is in clear contrast to the CP force of the same molecule placed near a metal plate where the behavior of the CP force depends crucially on the molecular polarization, on the  sign of $\alpha_{\parallel}-2\alpha_{z}$, to be specific, and the transition of attractive-to-repulsive is possible for molecules both in the ground and excited states~\cite{Z-Y12}.  The behaviors of the CP force are also sharply different when the molecule-substrate separation is comparable to the transition wave-length and the most outstanding feature is that the CP force for isotropically polarizable molecules can change from attractive to repulsive beyond a certain threshold temperature, which is about 320K for  a granite substrate. However, for the same molecule near a metal surface, the CP force is always attractive and this kind of changeover never happens.

For the case of  being out of thermal equilibrium but in  a stationary regime,  when the molecule-substrate separation is much less than the molecular transition wave-length, the CP force in the high temperature limit can be dramatically manipulated  by varying  the relative magnitude of the  temperatures of the substrate and the environment.  In particular, the attractive-to-repulsive transition can occur beyond a certain threshold temperature of either the substrate  or the environment  for molecules both in the ground and excited states.  If the temperature of the substrate is much higher, then there exists a threshold for the substrate temperature beyond which the transition happens for the molecules
in the ground state while the CP force for the excited states is always attractive. And it is just the other way around if the environment is of a much
higher temperature, i.e., the CP force for the ground state is attractive while that for the excited state can change from attractive to repulsive beyond
some threshold temperature of the environment. On the other hand, when the separation is comparable to the wave-length, the attractive-to-repulsive transition  which is impossible  when the substrate and the environment are at thermal equilibrium, may happen for molecules polarizable along the surface of the substrate with a small permittivity, while for isotropically polarizable ones the transitions can occur  even at room temperature for some dielectric substrate such as sapphire and graphite, and the transition temperature is much lower than the temperature for the same transition to take place in the thermal equilibrium case, thus making the
experimental demonstration of such force transitions easier.

\begin{acknowledgments}

This work was supported in part by the NSFC under Grants No. 11375092, No. 11435006 and No. 11405091,  the SRFDP under Grant No. 20124306110001, the Zhejiang Provincial Natural Science Foundation of China under Grant No. LQ14A050001, the Research Program of Ningbo University under Grants No. E00829134702, No. xkzwl10 and No. XYL14029, and the K.C. Wong Magna Fund in Ningbo University.

\end{acknowledgments}

\end{document}